\documentclass[conference]{IEEEtran}
\ifCLASSINFOpdf
  \usepackage[pdftex]{graphicx}
  \graphicspath{ {./assets/} }
\else
\fi
\begin{document}

\title{Vulnerability Analysis of Time Synchronization in Automotive Ethernet }

\author{\IEEEauthorblockN{Rishikesh Kakade$^{*1}$, Joey Chou$^{*2}$, Shannon Torcato$^{*3}$}
\IEEEauthorblockA{$^*$Department of Electrical and Computer Engineering, University of Waterloo\\
\{$^1$rkakade, $^2$t6chou, $^3$storcato\}@uwaterloo.ca\\}}


%


\maketitle

\begin{abstract}
The operation of many network communication protocols require accurate time synchronization between nodes. In the automotive space, IEEE 802.3bw (commonly referred to as automotive ethernet) is quickly becoming the most popular in-vehicle communication protocol between electronic control units (ECUs). The rapid advance of autonomous vehicles is predicated on a high throughput of multiple HD video streams, LIDAR readings, ML inferences, and vehicular control signals propagating on the same bus. To ensure reliability, security, and safety, for millions of passengers and bystanders, it is essential that the stack is robust at even the lowest level. The time synchronization protocol in this standard is IEEE 802.1 AS which was initially developed for virtual and virtual bridge local area networks (LANs). This paper presents an analysis of failure modes affecting the ability of in-vehicle automotive ethernet networks to synchronize clocks between ECUs. To demonstrate this, we use the OMNet++ discrete event simulator to showcase normal operation,  failover to the hot-standby clock, and a black hole attack on one network switch. 
\end{abstract}


%
\IEEEpeerreviewmaketitle

\section{Introduction}
In the early 1980s, automotive manufacturers began including electronic control units inside their cars. These tiny devices were the first step away from manual, mechanical control of cars by users, toward increased automation. in 1983, Bosch introduced the Controller Area Network protocol that arbitrated communication between ECUs. The protocol synchronizes between nodes using a synchronization byte that is transmitted at the start of each data packet transmitted. In the present day, this progression has led to an average car containing over 100 discrete ECUs, each responsible for controlling the millions of tiny electromechanical adjustments that make driving as effortless as it is today. 

The next evolution of this rapid progress is the the autonomous vehicle space. The amount of data a vehicle generates and processes has increased drastically due to an increase in the number of sensors present in the vehicle, some of which are LiDAR, cameras, and radar. These sensors are used to scan the surroundings of the vehicle to detect other vehicles and road conditions in real time. Due to this increase in data throughput required inside a vehicle, car manufacturers have reached the limits of the measly 2-wire, 500kbps,  CAN protocol. 

The leading candidate to succeed the CAN protocol is ethernet. Ethernet has data throughput rates of 100 Mbps --- representing a 200x improvement in transmission rate. It is a natural choice because the concept has been well tested and proven as the interface that connects the the internet. The large body of knowledge already available enables the industry to quickly adopt it and integrate into existing systems. 

A foundational, yet often overlooked component of the ethernet standard in this application is its method of synchronizing time across ECUs. It leverages Generic Precision Time Protocol (gPTP). The protocol synchronizes by comparing the difference in time between 2 master clocks. All other slave nodes synchronize to the master nodes by listening periodically for synchronization pulses transmitted from the master clock to many slaves clocks. As a redundancy measure, the protocol allows slaves to synchronize to more than one master clock. This acts as a fail safe in case one master clock loses its link to the rest of the network.

This method of synchronization is robust enough for most internet applications. However in the automotive space, every component in the car is subject to an incredible amount of scrutiny because of the enormous potential for loss of life. The time synchronization layer provides room for a number of failure modes that could cause critical loss of function and possibly, of life. Additional risk is introduced with the fact that modern vehicles are exposed to the outside world over the internet. From any layer of the stack, external adversaries could carry out an attack that exploits a vulnerability.

TSN which stands for Time-Sensitive Networking is a protocol defined by IEEE 802 standards that by default makes Ethernet deterministic. In particular, TSN addresses the transmission which are bounded by low latency and which has high availability. Some of the real-world applications are automotive vehicles or industrial control facilities that use real-time control streams.

Attacks could include jamming, routing infrastructure attacks, and byzantine attacks. With this large attack surface, it is especially important that designers keep robustness in mind. By simulating this time synchronization protocol we are able to perform experimentation at a lower cost. When it comes to a global time clock, there is no one standard time that is followed. Each node has its own internal clock that it uses to keep a track of events. Over time the clocks in the network diverge due to finite clock accuracy.

This paper will investigate a number of failure modes Key metrics of interest are global convergence time, clock offset, clock drift, and the maximum number of adversarial faults that can be tolerated.
\newpage
\section{Literature Review}

In \cite{Kim2015GatewayFF}, the authors propose a new gateway framework based on the controller area network (CAN) for in-vehicle networks (IVNs). The proposed system is easy to reuse and verify this helps reduce the development costs and time. This system introduces new functionalities such as parallel programming, diagnostic routing, network management (NM), dynamic routing update, multiple routing configuration, and security.

In \cite{Tuohy2015IntraVehicle}, the authors describe why Ethernet is more efficient than other legacy technologies such as CAN and MOST. Ethernet is preferred due to its increased bandwidth. The sensors all around the vehicle help build new and exciting features for safety such as collision avoidance, lane departure detection, traffic sign classification, blind spot detection, driver intent detection, pedestrian detection, automatic cruise control, and many others as described in \cite{Doshi2011Multimodal}.

The authors of \cite{Zhao2018ComparisonOT}  compare different clock synchronization algorithms against traffic types with individual timing requirements. The authors were able to focus on strategies for scheduling flows and factors leading to delay. They were able to find that the bandwidth of the flow is reserved based demand of the actual bandwidth, this helps introduce a more stable transmission. When adopted alone SPQ ruler can lead to high-priority data with large flow bursts which would lead to low priority flow to have high delays.

In the paper \cite{clocksyn}, the authors present a novel approach in PTP systems to reduce the clock time drifting significantly called timing fault recovery (TFR). In this approach, the fault occurrence is detected by TFR and within a short duration, it is recovered using a handshake mechanism. With increased reliability and availability of PTP systems, the TFR approach also provides constancy and clock stability.

The authors of \cite{ieeeClksyn} proposed and designed an IEEE 1588 protocol for a clock synchronization system using FPGA. This system is very accurate and can detect time precision at the microsecond level in the synchronization of the master clock and slave clock. This design is an improvement to the current time synchronization precision, it also helped reduce the complex hardware circuit design, and helped reduce the development cost and difficulty, this provides a benchmark value to research done in the future.

In \cite{can}, the authors discuss the performance of an autonomous vehicle. The data measured from various sources are integrated and processed, depending on whether time information is available from the sensor measurement the data will vary. Temporal distortion of the controller in which missing patches of information are patched up at different instants so the situation can be perceived. On the other hand, time information-based sensor measurements are aware of the time differences among the values this allows the autonomous driving controller to have a better understanding of the situation by compensating for the time differences. The authors conclude that to increase autonomous vehicle performance, time synchronization-based research should be looked into thoroughly sensors and controllers connected via communication networks.

\cite{timesyncgptp} presents a very similar analysis of the weaknesses of gPTP as is presented in this paper. Man-in-the-middle (MIM) attacks operate in a number of ways such as by intercepting and removing valid synchronization messages, by manipulating messages, and by delaying legitimate messages. Denial of Service (DoS) attacks often work by flooding the network with messages. Time source attacks involve corrupting the source of truth the master clocks rely on (GPS fraud) one of the master clocks being compromised. Master selection attacks operate by having an attacking node send the same messages a master node would, fooling slave nodes to synchronize to the wrong clock. The report also suggests counter measures against these attacks. 

In \cite{Puttnies2018ASM}, the authors created a simulation model of the IEEE 802.11AS generalized Precision Time Protocol(gPTP) standard on OMNET++ for TSN specific synchronization. The authors utilize measurements of propagation delay and time synchronization between systems that are time-aware. The model described in this papers utilizes clocks with constant drift and end stations. The results observed that every one second propagation delay is measured whereas it was every 125 ms or 62.5 ms for synchronization message sent from GM. As a result of this there is a difference in clock drifts. The time difference of all the nodes after synchronization to GM is zero as expected from the gPTP simulation model.

\textbf{Black Hole Attack} 
In this type of attack, the adversary stops propagating forward the information or data but is still an active node in the routing protocol.
The adversarial node when part of a routing protocol path prevents further communication to take place on that part. This type of attack affects secure and insecure routing protocols due to their nature of 
rendering the normal methods of route maintenance useless

\textbf{Flood Rushing Attack} 
In this attack, the flood duplicate suppression technique of the node is exploited. The adversarial flood tries to be the first to affect the nodes over the actual flood during propagation. When a node receives a flood packet from an adversary then the nodes will ignore the legitimate version and will forward propagate the adversarial version. Due to this, it is almost impossible for the node network to be free from adversarial propagation.

\textbf{Byzantine Wormhole Attack}
In this kind of attack, the compromised nodes communicate with each other in order to gain an advantage. By communicating with each other they are able to create a shortcut in the network. By sending out a route request an adversary can discover a route in the network, to execute the attack, the tunnel packets through the non-adversarial nodes. In order to increase the probability of being selected as the route path, the adversaries use the low cost appearance of the wormhole links. They then drop all of the data packets to disrupt the network. By having as minimum as two compromised nodes this particular attack can be performed.

\textbf{Byzantine Overlay Network Wormhole Attack}
This kind of attack is similar to the wormhole attack but in this case, several nodes are compromised and they form a network on top of the existing network. The adversaries tunnel packets through the overlay network and make it seem like they are all neighbours to the routing protocol. This helps increase the chances of being selected on routes [1].

\section{Methodology}

To build our model we use an OMNet++ simulation with the INET 4.4 library \cite{inet1}. 

We can combine the gPTP model with the INET library along with the other standard network protocols that are present in the library. For the simulation model, we do not consider the best master clock algorithm of the gPTP protocol. Therefore, we predefine the state of the node as master or slave. We are more interested in clock synchronization and measurement of propagation delay. We introduce a clock drift so that we can implement a time synchronization protocol. 

In order to meet gPTP requirements, we implement time synchronization and propagation delay measurements using a class from the OMNeT library \cite{inet2}. We use the in-built EthernetIIFrame available in the INET library which has encapsulated gPTP messages.

In INET, network nodes and interfaces do not have local time. Instead, the global time is defined as the simulation time. Network clock submodules are required by network nodes to simulate local time-based network nodes and to demonstrate effects such as clock drift. The oscillator submodules are used to keep a track of time. The basic working is that the oscillator periodically produces ticks which are counted by the clock modules. Inaccuracies in oscillators can lead to clock drift.

There can be drift in clocks of the time kept between different network devices. When two clocks are synchronized, the time difference between them has an upper bound. We can achieve this between clocks by periodically synchronizing them. In gPTP, there is a time synchronization of the slave clocks to the time of the master clocks within a time domain of the gPTP. Multiple time domains of gPTP can be obtained by a network. Multiple clocks can be synchronized to multiple master clocks in case there is a failure in the master clock or a link break causes the system to go offline. The general structure is that there exists a single master clock followed by a number of slave clocks. Synchronization by the protocol is achieved by sending sync messages to the slave clocks from the master clock. By having multiple gPTP submodules, a node can have multiple gPTP time domains. Depending on their location, a gPTP can follow one out of three possible roles: master, bridge, or slave. By having multiple gPTP submodules, a node can have multiple gPTP time domains. Depending on their location, a gPTP can follow one out of three possible roles: master, bridge, or slave. Master gPTP modules of nodes contain the time domain master clock.

\begin{figure}[!t]
\centering
\includegraphics[width=3.5in]{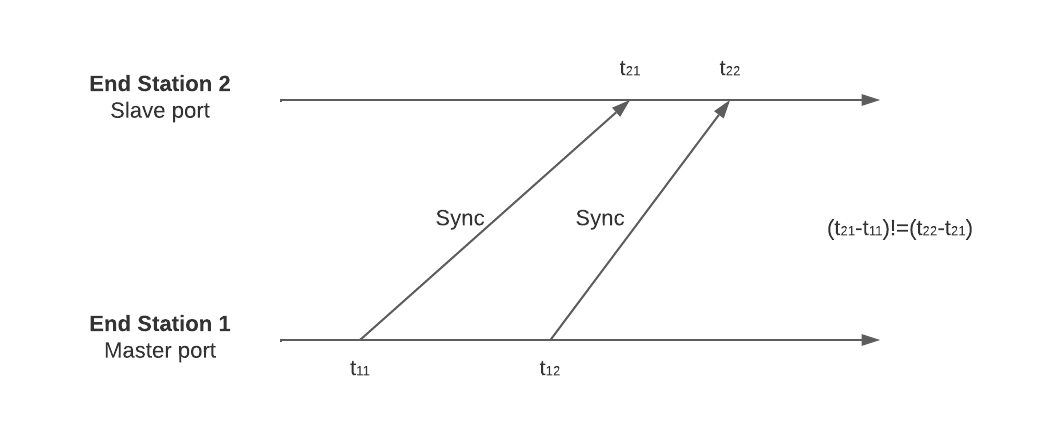}
\caption{Arrival of messages due to impact of clock drift on gPTP network.}
\label{drift}
\end{figure}

\begin{figure}[!t]
\centering
\includegraphics[width=3.5in]{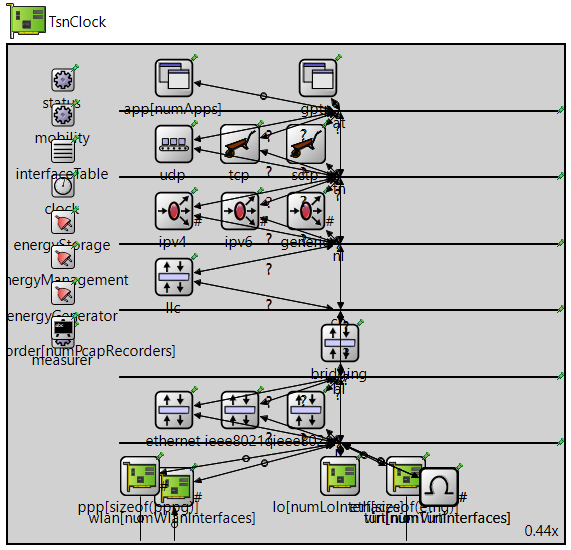}
\caption{TSN Clock model taken from iNet class library. }
\label{tsn}
\end{figure}

The fundamental device underlying the simulation the clock seen in Figure \ref{tsn}. It is modelled using basic clock classes from the INET library \cite{inet1}. The basic models include the ability to represent clock drift. Clock drift arises due to inaccuracies in the low-cost crystal oscillators often used in automotive ECUs. In the simulation, we have modelled the  drift as a random walk, with steps varying by $\pm 100\mu$s every second. In the real world, this variance is generally caused by process variations, thermal expansion, and device aging. The change in clock drift over time is ignored because it changes very slowly relative to the time frames on which automotive ethernet networks operate. For this short simulation, clock frequency is assumed to be constant.

\begin{figure}[!t]
\centering
\includegraphics[width=3.5in]{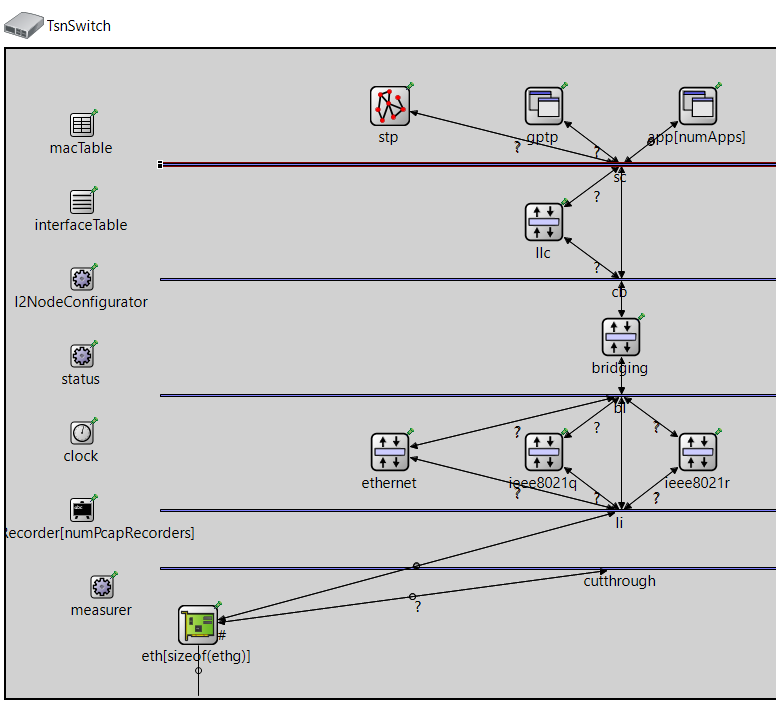}
\caption{TSN Switch model taken from iNet class library.}
\end{figure}

\begin{figure}[!t]
\centering
\includegraphics[width=3.5in]{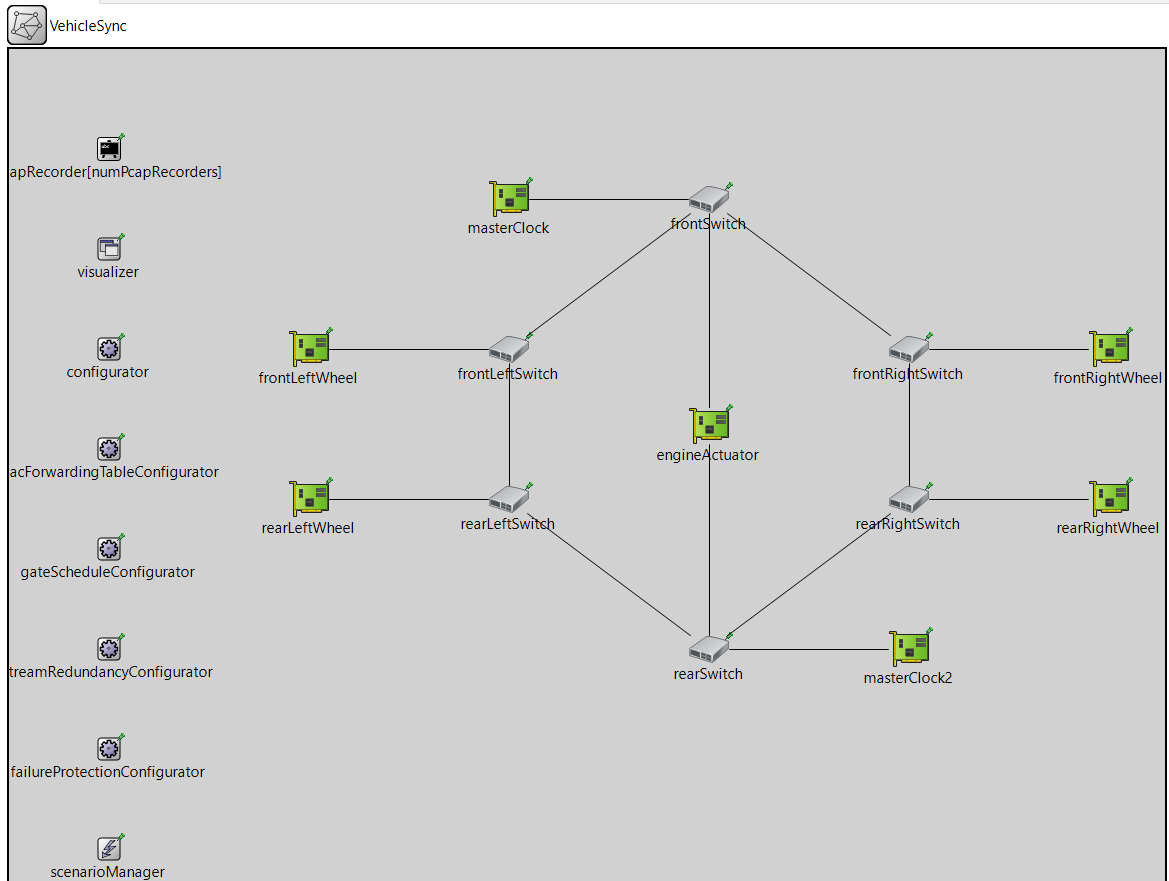}
\caption{Wheel motor controller network whose switches are in a ring topology. Contains double redundant clocks representing central car computer and central motor body controller.}
\label{ned}
\end{figure}

We emulate 4 ECUs controlling the 4 DC motors of an electric car. The ECUs are synchronized to a central body controller and the main computer as a redundancy measure. In the simulation the network is set up with two master clocks; each disseminating two time domains. Each node in the network keeps track of 4 total times. The switching network is shaped as a ring to further emphasize the benefit of having multiple time domains propagating in different directions as a redundancy measure.

To summarize the way the network in Figure \ref{ned} works \cite{inet1};
\begin{enumerate}
    \item Primary clock has one clock, two master time domains. The domains are propagated in both clockwise and counterclockwise directions along the ring
    \item Secondary clock has 4 time domains. Two are slaves to the primary clock's domains while the others are redundant domains. 
    \item all other nodes in the network are slaves to the four total time domains propagating in the network
\end{enumerate}

This paper presents three scenarios. The first is operation of the network under normal conditions. Under normal circumstances, the only disturbance that needs to be corrected for is clock drift as described in Figure \ref{drift}. It shows that between successive synchronization pulses from the master nodes to each slave, the relative clock drift causes the delta between each local time to be non-constant. In the second scenario, we simulate a failure of the central body controller. This could be caused by a number of failure modes, such as power loss, bus message overloads, and firmware failures. We expect that the slaves in the network should be able to switch over to the redundant secondary master in the car's central computer. In the third, we simulate a black hole attack on one of the network switches. Here, we make the switch unable to forward synchronization messages to the slave ECU it is connected to. It is expected that the device should be unable to synchronize with the rest of the network. 

The relevant simulation files are made open source on Github \cite{git}.

\section{Results}
\subsection{Normal operation}
\begin{figure}[!t]
\centering
\includegraphics[width=3.5in]{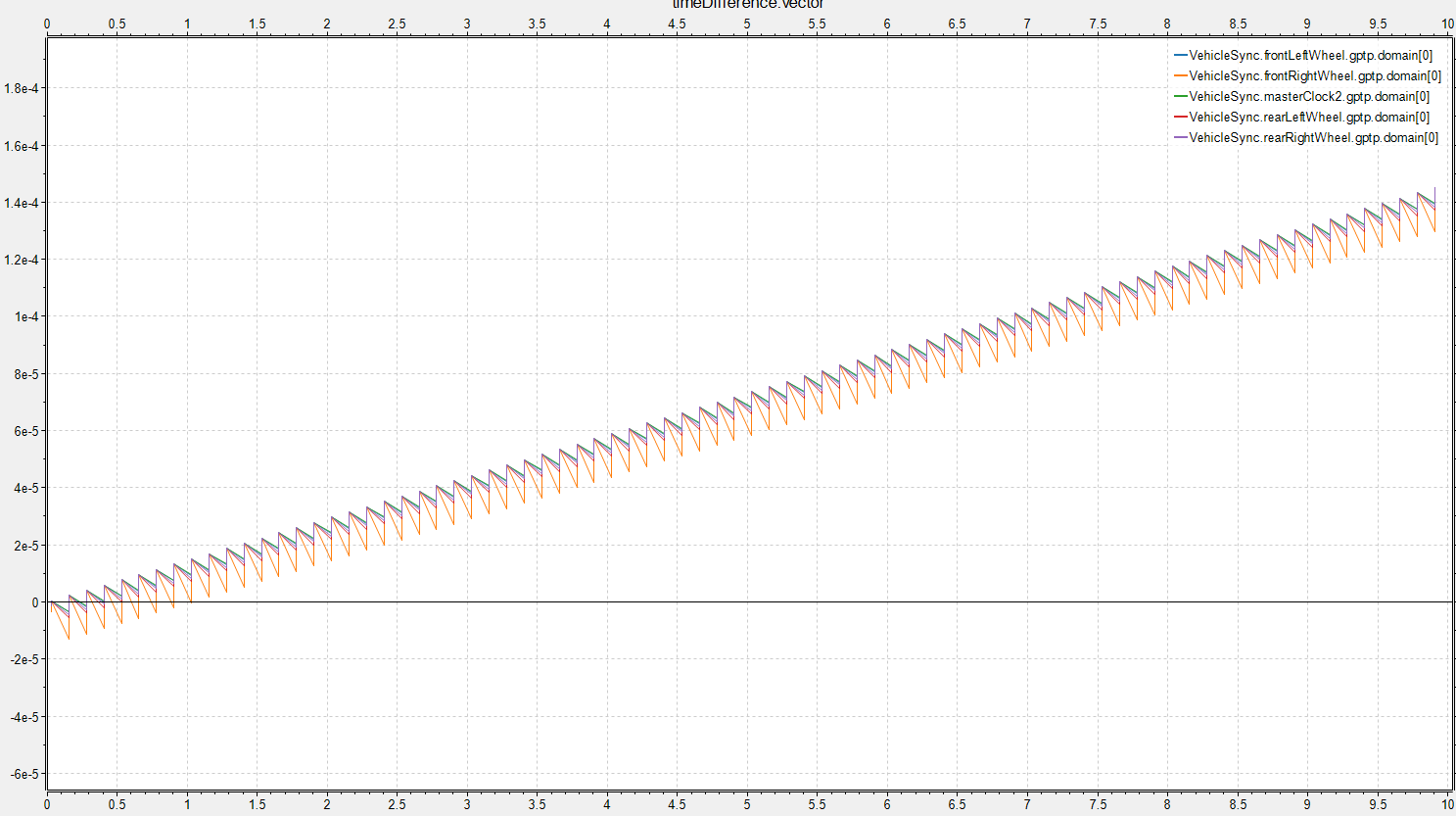}
\caption{Difference between device time and true simulation time under normal conditions of the domain disseminated by the primary clock in the body controller.}
\label{normal}
\end{figure}

Figure \ref{normal} shows the difference between the clocks from 'true' simulation time of the network under normal conditions. The clock in the front right wheel has the most dramatic clock drift which explains why peak to peak difference is higher than in others. Despite this, the network is still able to synchronize all components to within about $\pm 100$ns of one another. We include simulation of the other, redundant time domain disseminated by the main computer at this time. We find that all nodes in the network are able to synchronize to this as well, even though it has a different clock drift. When the end-to-end delay is constant there is no clock drift as the generated packets are in sync with the source. 

\begin{figure}[!t]
\centering
\includegraphics[width=3.5in]{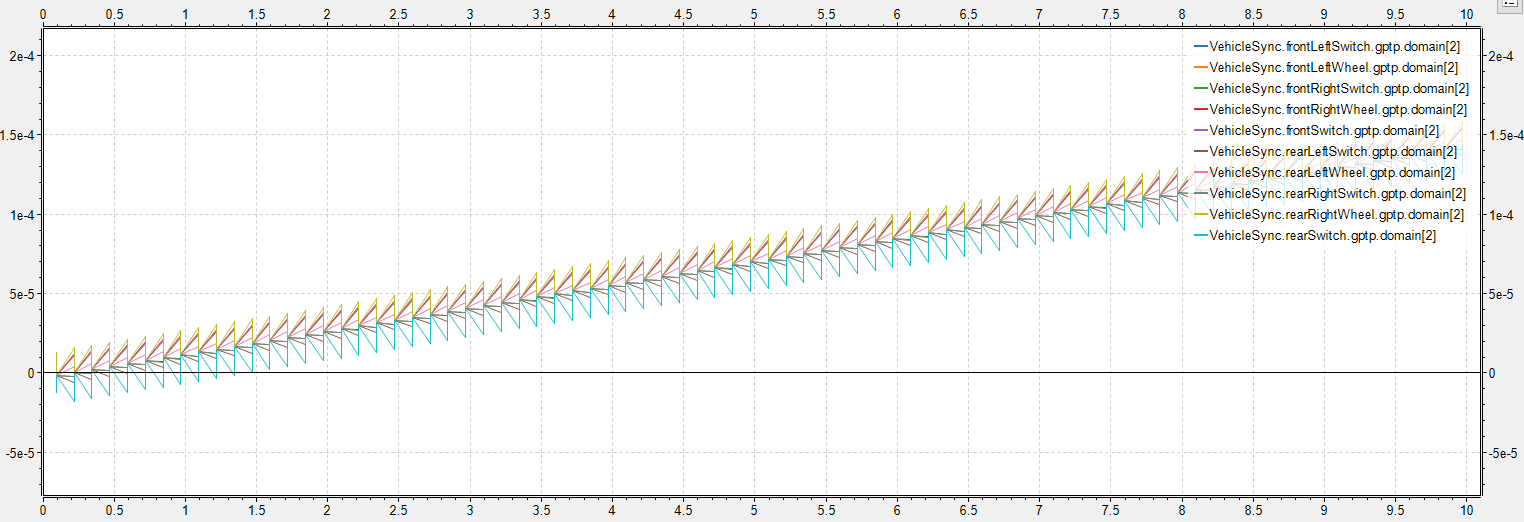}
\caption{Difference between device time and true simulation time under normal conditions disseminated by the secondary clock in the main computer. Observe that the drift rate is different between this and Figure \ref{normal} because it originates from a different clock. }
\label{normal2}
\end{figure}

\subsection{Failure of primary clock}
\begin{figure}[!t]
\centering
\includegraphics[width=3.5in]{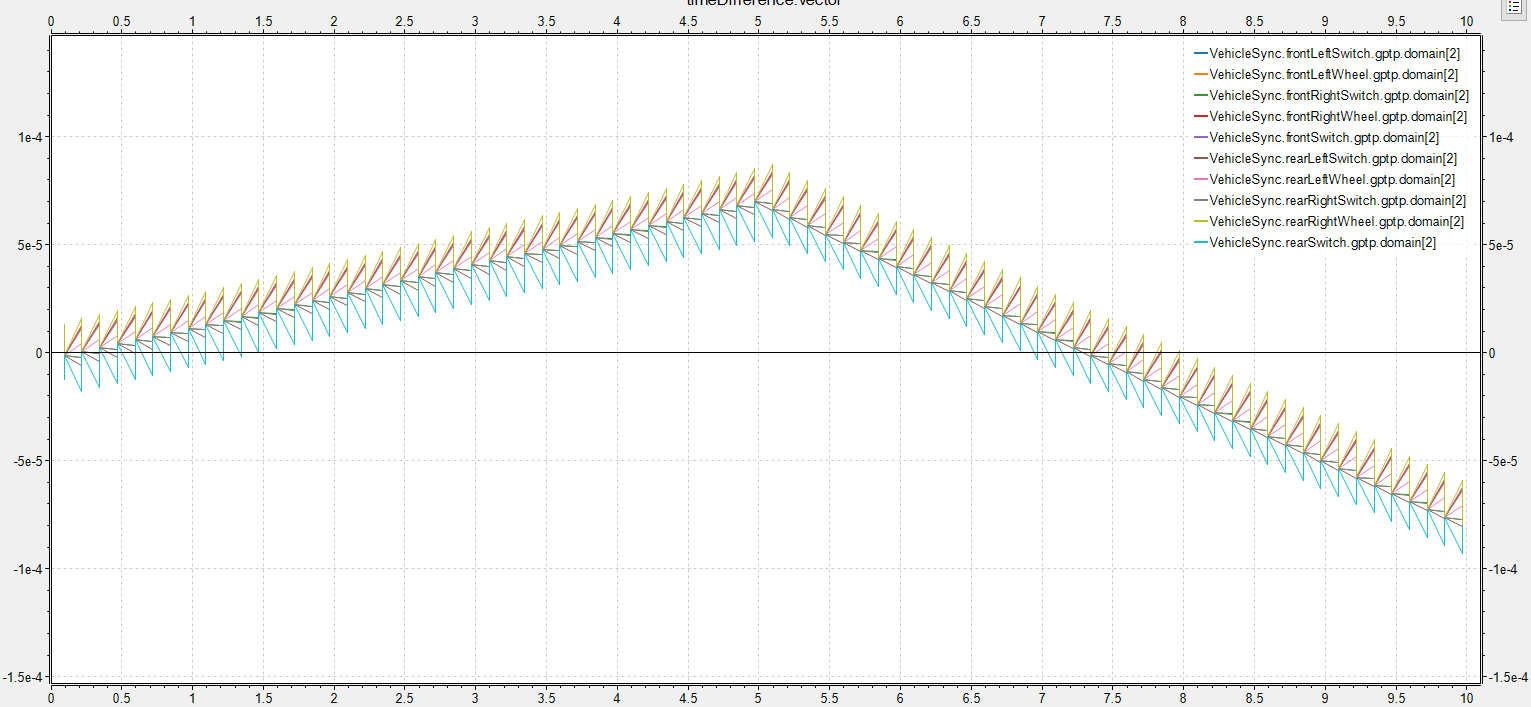}
\caption{Difference between device time and true simulation time when the central body controller (primary clock) fails for domain 2 originating in the central computer (secondary). Observe that the system is able to switch over to the main computer clock after the body controller fails.}
\label{clk_fail}
\end{figure}

Figure \ref{clk_fail} shows the difference between the clocks from 'true' simulation time of the network when the primary clock in the body controller fails. Unlike in the normal mode simulation, we find that as soon as the primary clock fails, all time domains disseminated by it cease to exist. We find that the simulator does not simulate past this event where a domain stops existing. However, the overall system is able to switch over to time domain 2 which is disseminated by the secondary clock in the main computer. The primary master clock in this configuration has a link failure as a result there is a stoppage in the two time domains' time synchronization of the primary master clock. After our master clock fails, the active clocks differ in their time stamp.

\begin{figure}[!t]
\centering
\includegraphics[width=3.5in]{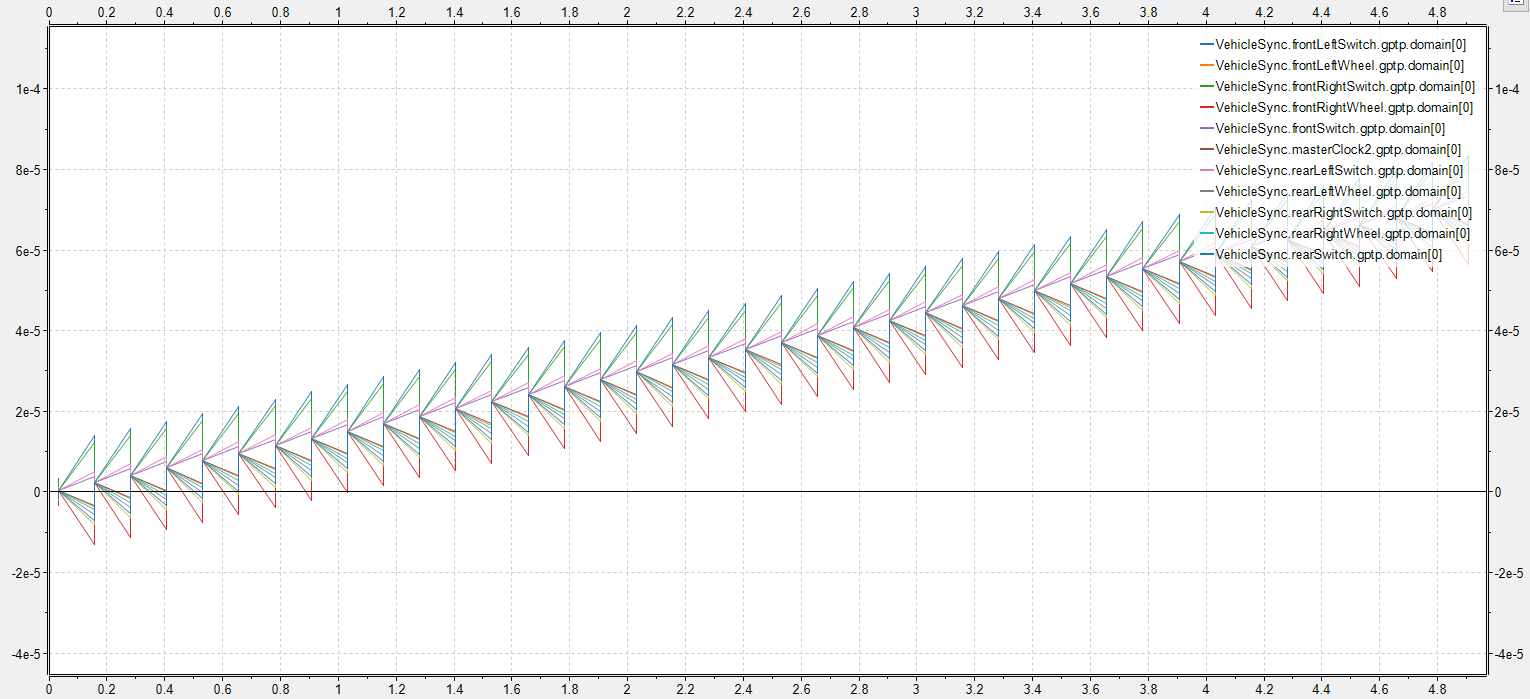}
\caption{Difference between device time and true simulation time when the central body controller (primary clock) fails for domain 0 originating in primary clock. Observe that the simulation stops at the moment the primary node fails.}
\label{failover}
\end{figure}

\subsection{Black hole network switch}
\begin{figure}[!t]
\centering
\includegraphics[width=3.5in]{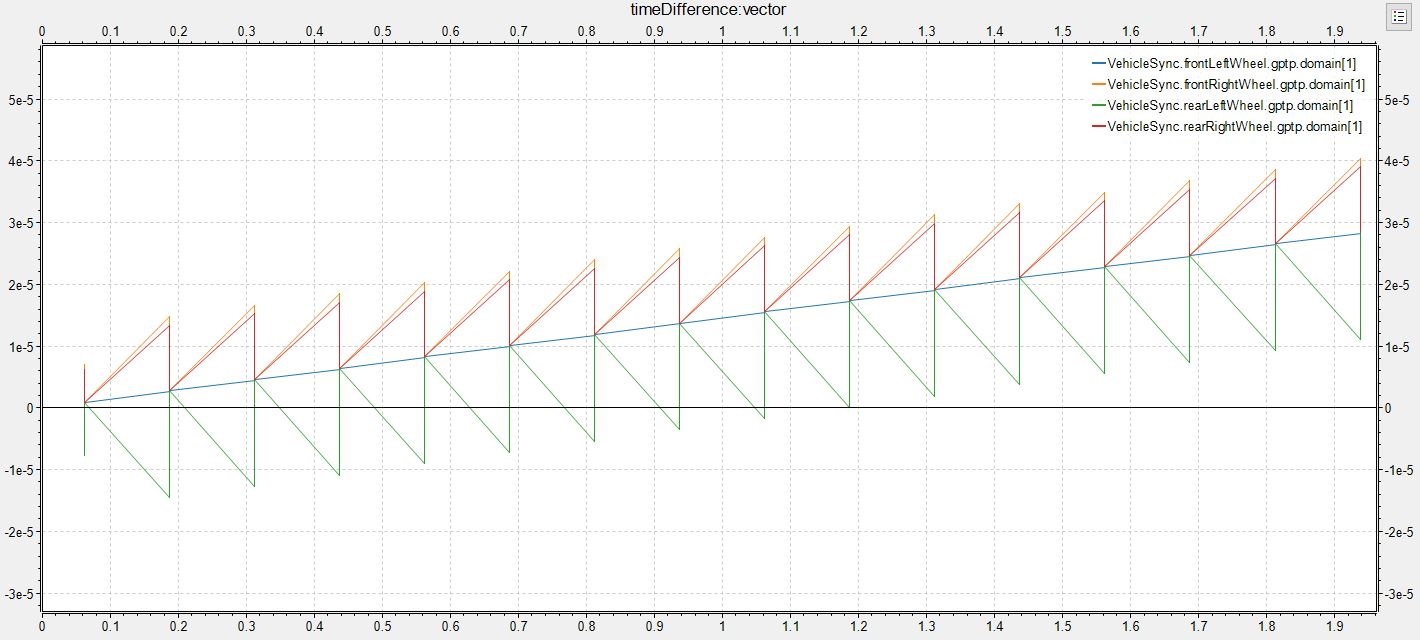}
\caption{Difference between device time and true simulation time when the front-left switch refuses to forward synchronization messages to the front-left motor ECU. Observe that the ECU is unable to synchronize with the rest of the network and just ticks at its own clock drift. }
\label{bing}
\end{figure}

Figure \ref{bing} shows the difference between the clocks from 'true' simulation time of the network when a black hole attack is performed on the front-left network switch which forces the front-left motor ECU to be left nonoperational. We see that this is the case by the thin blue line in Figure \ref{bing}. The clock drift of this ECU was preset to be the same as the master clock to create a simpler graph. 

\section{Conclusion}

In this paper, we have proposed a gPTP simulation model containing the main operations: time synchronization and propagation delay measurements. We demonstrated the failure of different modes affecting the ability of synchronization of ethernet networks present in automotive vehicles between ECUs. We used OMNet++ as the simulation tool to showcase different scenarios such as normal operation, fail-over to the hot-standby clock, and a black hole attack on one network switch. We also showed how a black hole attack can effect the time synchronization between master clock and slave clocks. Through our experimentation, we demonstrated the vulnerabilities that exist in the present ethernet networks present in automotive vehicles. Further research can be done to overcome this vulnerability and create a more secure ethernet network system by implementing additional redundancy checks that each ECU node is synchronized with the rest of the network.

\section*{Acknowledgment}
The authors would like to thank Prof. Otman Basir for giving the time and resources to choose and develop our ideas into this report.



%

\nocite{*}
\bibliographystyle{IEEEannot}
\bibliography{refs}

\begin{thebibliography}{10}
\providecommand{\url}[1]{#1}
\csname url@rmstyle\endcsname
\providecommand{\newblock}{\relax}
\providecommand{\bibinfo}[2]{#2}
\providecommand\BIBentrySTDinterwordspacing{\spaceskip=0pt\relax}
\providecommand\BIBentryALTinterwordstretchfactor{4}
\providecommand\BIBentryALTinterwordspacing{\spaceskip=\fontdimen2\font plus
\BIBentryALTinterwordstretchfactor\fontdimen3\font minus
  \fontdimen4\font\relax}
\providecommand\BIBforeignlanguage[2]{{%
\expandafter\ifx\csname l@#1\endcsname\relax
\typeout{** WARNING: IEEEtran.bst: No hyphenation pattern has been}%
\typeout{** loaded for the language `#1'. Using the pattern for}%
\typeout{** the default language instead.}%
\else
\language=\csname l@#1\endcsname
\fi
#2}}

\bibitem{timesyncgptp}
\BIBentryALTinterwordspacing
 [Online]. Available:
  \url{https://standards.ieee.org/wp-content/uploads/import/documents/other/d1-10\_jesse\_making\_gptp\_capable\_for\_secure\_time\_synchronization.pdf}
\BIBentrySTDinterwordspacing


\bibitem{spec}
\BIBentryALTinterwordspacing
``802.1as - timing and synchronization.'' [Online]. Available:
  \url{https://www.ieee802.org/1/pages/802.1as.html}
\BIBentrySTDinterwordspacing


\bibitem{inet1}
\BIBentryALTinterwordspacing
``Effects of time synchronization on time-aware shaping.'' [Online]. Available:
  \url{https://inet.omnetpp.org/docs/showcases/tsn/combiningfeatures/
  gptpandtas/doc/index.html}
\BIBentrySTDinterwordspacing


\bibitem{inet2}
\BIBentryALTinterwordspacing
``Using gptp¶.'' [Online]. Available:
  \url{https://inet.omnetpp.org/docs/showcases/tsn/timesynchronization/gptp/
  doc/index.html}
\BIBentrySTDinterwordspacing


\bibitem{git}
\BIBentryALTinterwordspacing
``Vulnerability-analysis-of-time-synchronization-in-automotive-ethernet.''
  [Online]. Available:
  \url{https://github.com/RishiKakade/Vulnerability-Analysis-of-Time-Synchronization-in-Automotive-Ethernet}
\BIBentrySTDinterwordspacing


\bibitem{autoether}
\BIBentryALTinterwordspacing
B.~Bob and L.~Lo, ``The case for ethernet in automotive communications,''
  \emph{SIGBED Rev.}, vol.~8, no.~4, p. 7–15, dec 2011. [Online]. Available:
  \url{https://doi.org/10.1145/2095256.2095257}
\BIBentrySTDinterwordspacing


\bibitem{Doshi2011Multimodal}
A.~Doshi, B.~T. Morris, and M.~M. Trivedi, ``On-road prediction of driver’s
  intent with multimodal sensory cues,'' \emph{IEEE Pervasive Computer},
  vol.~10, p. 22–34, 2011.


\bibitem{can}
\BIBentryALTinterwordspacing
H.~J. Kim, U.~Lee, M.~Kim, and S.~Lee, ``Time-synchronization method for
  can–ethernet networks with gateways,'' \emph{Applied Sciences}, vol.~10,
  no.~24, p. 8873, Dec 2020. [Online]. Available:
  \url{http://dx.doi.org/10.3390/app10248873}
\BIBentrySTDinterwordspacing


\bibitem{Kim2015GatewayFF}
J.~Kim, S.-H. Seo, N.~T. Hai, B.~M. Cheon, Y.~S. Lee, and J.~W. Jeon, ``Gateway
  framework for in-vehicle networks based on can, flexray, and ethernet,''
  \emph{IEEE Transactions on Vehicular Technology}, vol.~64, pp. 4472--4486,
  2015.


\bibitem{clocksyn}
\BIBentryALTinterwordspacing
A.~A. Omar and J.~M. Rhee, ``Tfr: A novel approach for clock synchronization
  fault recovery in precision time protocol (ptp) networks,'' \emph{Applied
  Sciences}, vol.~8, no.~1, 2018. [Online]. Available:
  \url{https://www.mdpi.com/2076-3417/8/1/21}
\BIBentrySTDinterwordspacing


\bibitem{Puttnies2018ASM}
H.~Puttnies, P.~Danielis, E.~Janchivnyambuu, and D.~Timmermann, ``A simulation
  model of ieee 802.1as gptp for clock synchronization in omnet++,'' 2018.


\bibitem{Tuohy2015IntraVehicle}
S.~Tuohy, M.~Glavin, C.~Hughes, E.~Jones, M.~Trivedi, and L.~Kilmartin,
  ``Intra-vehicle networks: A review,'' \emph{IEEE Transactions on Intelligent
  Transportation Systems}, vol.~2, pp. 534--554, 2015.


\bibitem{ieeeClksyn}
B.~Zhao and N.~Wang, ``The implementation of ieee 1588 clock synchronization
  system based on fpga,'' pp. 216--220, 2014.


\bibitem{Zhao2018ComparisonOT}
L.~Zhao, F.~He, E.~Li, and J.~Lu, ``Comparison of time sensitive networking
  (tsn) and ttethernet,'' \emph{2018 IEEE/AIAA 37th Digital Avionics Systems
  Conference (DASC)}, pp. 1--7, 2018.


\end{thebibliography}

\end{document}